\documentclass{svjour3}   
\smartqed  
\usepackage{graphicx}
\usepackage{pxfonts}
\journalname{General Relativity and Gravitation}

\begin{document}
\title{Modeling the motion of a bright spot in jets from black holes M87* and SgrA*}
\author{Vyacheslav I. Dokuchaev 
	\and Natalia O. Nazarova}
\institute{V. I. Dokuchaev \at Institute for Nuclear Research of the Russian Academy of Sciences, 
60th October Anniversary Prospect 7a, 117312 Moscow, Russia, \\\email{dokuchaev@inr.ac.ru} 
\and
N. O. Nazarova \at Scuola Internazionale Superiore di Studi Avanzati (SISSA), Via Bonomea 265, 34136 Trieste (TS) Italy \at
International Centre for Theoretical Physics (ICTP), Strada Costiera 11, 34151 Trieste (TS), Italy \\\email{natalia.nazarova@sissa.it}  }
\date{Received: date / Accepted: date}
\maketitle

\begin{abstract}
We study the general relativistic motion of a bright spot in a jet from an accreting black hole. The corresponding lensed images of the moving bright spot are calculated numerically in discrete time intervals along the bright spot trajectory in the Kerr space-time framework. As representative examples, we consider the cases of supermassive black holes SgrA* and M87*. Astrophysical observations of the moving bright spots in the jets from black holes provides the unique possibility for the verification of different gravitation theories in the strong field limit.
\end{abstract}

\section{Introduction}
\label{intro}

One of the most impressive manifestations of the Active Galactic Nuclei (AGN) is the powerful relativistic jets from their central enigmatic supermassive black holes \cite{Rees78,Rees78b,Eichler83,Rees84,Begelman84,Stiavelli92,Junor95,Junor99,Matteo03,Kovalev07,Hada11,deGasperin12,Moscibrodzka16,Doeleman12,Broderick15,Lacroix16,Akiyama17,Kawashima20,Mannheim00,Richards11,Kovalev12,Kovalev17,Kovalev19,Valverde20,Kovalev20a,Kovalev20,Kovalev20b,Kovalev20c,Troitsky20,Troitsky20b}. 

Nowadays, the most advanced models of these jets are studied with the help of the supercomputer General Relativistic Magneto-hydrodynamic (GRMHD) simulations \cite{Villiers05,Tchekhovskoy11,Tchekhovskoy12,Tchekhovskoy12b,Tchekhovskoy15,Tchekhovskoy17,Tchekhovskoy17b,Ryan18} based on the startling Blandford-Znajek process of energy extraction from the rotating black hole \cite{BlandfordZnajek,Beskin09,Beskin10,Toma16}.

The fast technical advance in astrophysical obser\-vations opens a new window for detailed investigations of the relativistic environs of astrophysical black holes. In particular, the crucial physical task is the observation and investigation of jets generated supermassive black holes at the centers of active galactic nuclei.

The gravitational lensing of any luminous object in the strong gravitational field of the black hole results in an infinite number of images  \cite{CunnBardeen72,CunnBardeen73,Viergutz93,RauchBlandf94,GralHolzWald19}. The brightest images (the so-called direct or prime images) are produced by the photons whose trajectories do not cross the equatorial plane of the rotating black hole. The secondary images (light echoes) or higher order images are generated by photons that intersect the black hole's equatorial plane several times. Note that the energy flux from all secondary images, as a rule, is negligible in comparison to one from the direct image. 

In this paper, we suppose that the motion of bright spot in the jet is ballistic. The corresponding trajectories of photons emitted by the moving bright spot and reaching a distant telescope (observer) are calculated using Carter equations of motion for test particles in the Kerr space-time. We calculate the positions, forms and brightness as for the direct lensed bright spot images and also for the two light echoes in discrete times along the trajectory of the bright spot in the jet.

In Appendix we collect some mathematical details and physical approaches for our numerical calculations of the bright spot motion along the black hole rotation axes.

\section{Kerr metric}

The classical form of the Kerr space-time \cite{Kerr,Chandra,BoyerLindquist,Carter68,deFelice,Bardeen70,Bardeen70b,BPT,mtw} is
\begin{equation}
	ds^2=-e^{2\nu}dt^2+e^{2\psi}(d\phi-\omega dt)^2
	+e^{2\mu_1}dr^2+e^{2\mu_2}d\theta^2,
	\label{metric}
\end{equation}
where
\begin{eqnarray}
	e^{2\nu}&=&\frac{\Sigma\Delta}{A}, \quad e^{2\psi}=\frac{A\sin^2\theta}{\Sigma}, 
	\quad \omega=\frac{2Mra}{A}, \\
	e^{2\mu_1}&=&\frac{\Sigma}{\Delta}, \quad e^{2\mu_2}=\Sigma, 
	\quad e^{2\mu_1}=\frac{\Sigma}{\Delta},  \\
	\Delta &= & r^2-2Mr+a^2, \quad 
	\Sigma = r^2+a^2\cos^2\theta,  \label{Sigma} \\
	A&=& (r^2+a^2)^2-a^2\Delta\sin^2\theta.  \label{A}
\end{eqnarray}
Here $M$ --- a black hole mass, $a=J/M$ --- a black hole spin. Throughout this paper we use the useful units: the Newtonian gravitational constant $G=1$ and the velocity of light $c=1$. Additionally, we use also the dimensional values (to simplify formulas): $r\Rightarrow r/M$, $t\Rightarrow t/M$ and similar ones. These mean that $GM/c^2$ --- is the used unit for the radial distance, and, correspondingly, $GM/c^3$ --- is the used unit for the time intervals. Similarly, $a=J/M^2\leq1$ (with $0\leq a\leq1$) --- is the dimensionless value of the black hole spin. The event horizon radius of the rotating black hole in these units is 
\begin{equation}
	r_{\rm h}=1+\sqrt{1-a^2}. \label{rh}
\end{equation}

\section{Moving hot spot in the jet from SgrA*}

The supermassive black hole Sagittarius A* (SgrA*) is in the center of our native Milky Way galaxy. A distant telescope (related with the planet Earth) locates near the black hole equatorial plane (at $\theta_0\simeq84^\circ\!\!.\,24$). The black hole equatorial plane coincides with the Galactic equatorial plane.

We calculate the trajectories of photons in the Kerr metric emitted by the bright blob of matter and reaching a distant telescope by using numerical solutions of integral equations  (\ref{rmot}--\ref{tmot}). In these calculations we suppose that black hole rotates with a maximal spin $a=1$.

\begin{figure}
	\includegraphics[angle=0,width=0.7\textwidth]{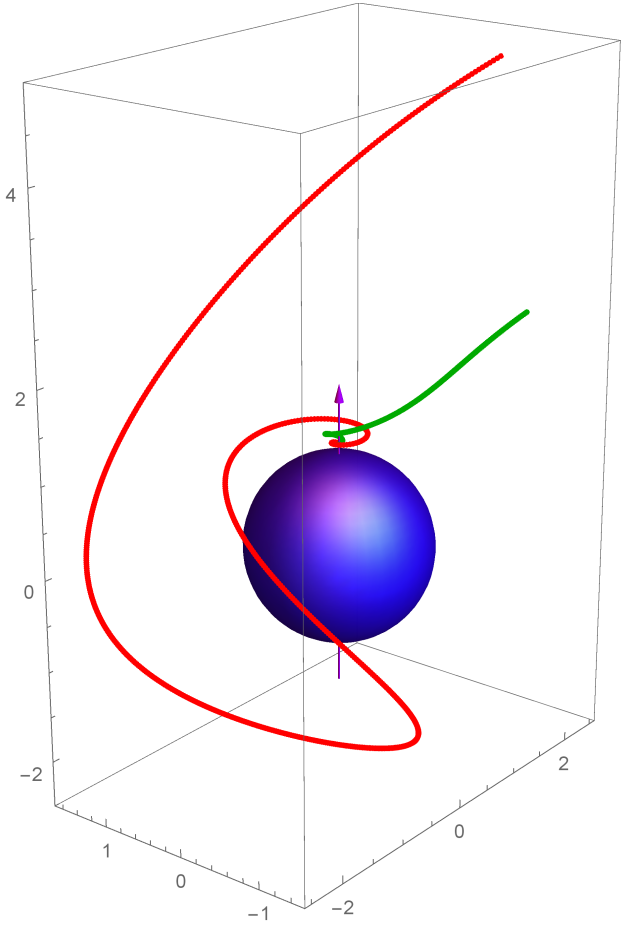}
	\caption{$3D$ trajectories of photons starting above the event horizon globe (blue sphere) from the radius $r=1.1r_{\rm h}$ on the rotation axis (marked by magenta arrow) and providing, correspondingly, the prime image (green curve) and the 1-st light echo (red curve) of the outward moving hot spot. The photon of the 1-st light echo moves clockwise in the $(r,\theta)$ plane, as clearly viewed in Fig.~\ref{fig2}.}	\label{fig1}
\end{figure}

Fig.~\ref{fig1} shows $3D$ trajectories of two photons starting above the event horizon globe from the radius $r=1.1r_{\rm h}$ on the rotation axis and providing the prime image (green curve) and the 1-st light echo (red curve) of the outward moving hot spot. The orbital parameters ($\lambda=0,\,q=1.75$) for the prime image photon and ($\lambda=0,\,q=4.665$) for the 1-st light echo photon (without the turning point $r_{\rm min}$ and moving clockwise in the $(r,\theta)$ plane) are derived from the numerical solutions of the integral equations of motion (\ref{eq24a}) and (\ref{eq24b}), respectively, with the photon starting point ($r_s=1.1,\theta_s=0$) and the final point ($r_0=\infty,\theta_0=84^\circ\!\!.\,24$).

\begin{figure}
	\includegraphics[width=0.5\textwidth]{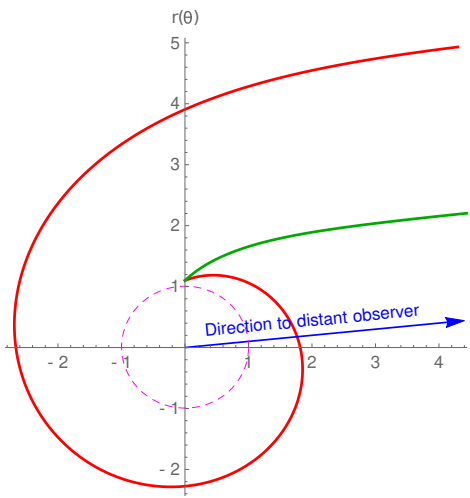}
	\caption{$2D$ version of Fig.~\ref{fig1}. Relations $r(\theta)$ for photons starting from radius $r=1.1r_{\rm h}$ on the rotation axis and producing a prime image (green curve) and, respectively, producing the 1-st light echo and moving clockwise in the $(r,\theta)$ plane (red curve). The magenta dashed circle is a position of the black hole event horizon with the radius $r_{\rm h}=1$ in the  Euclidean space without gravity.} \label{fig2}
\end{figure}

\begin{figure}
	\includegraphics[width=0.75\textwidth]{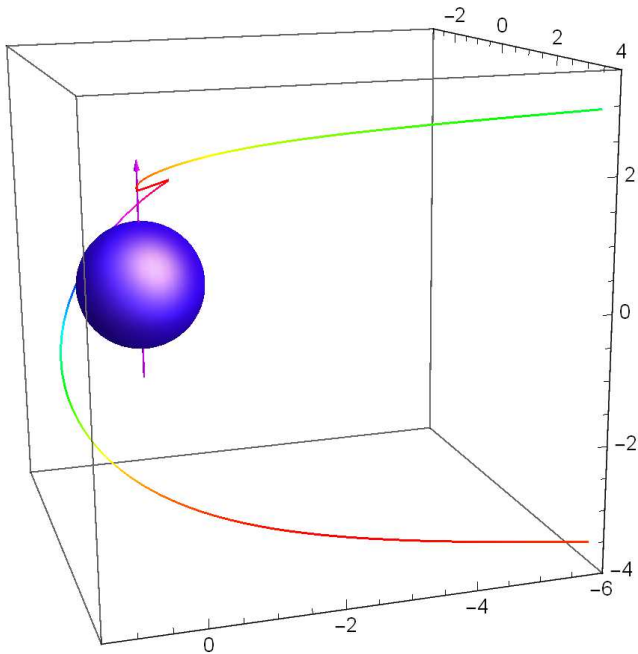}
	\caption{ $3D$ trajectories of photons, starting at the radius $r=1.49$ on the rotation axis of the black hole, and providing the prime image (upper multicolored curve) and the 1-st light echo (lower multicolored curve) of the outward moving hot spot. The orbital parameters of photons $\lambda=0$ and $q=2.175$ for the prime image and, respectively, $\lambda=0$ and $q=4.436$ for the 1-st light echo. The photon trajectory  of this 1-st light echo is without the turning point $r_{\rm min}$, and this photon is moving counterclockwise in the $(r,\theta)$ plane.} \label{fig3}
\end{figure}
\begin{figure}
	\includegraphics[angle=0,width=0.85\textwidth]{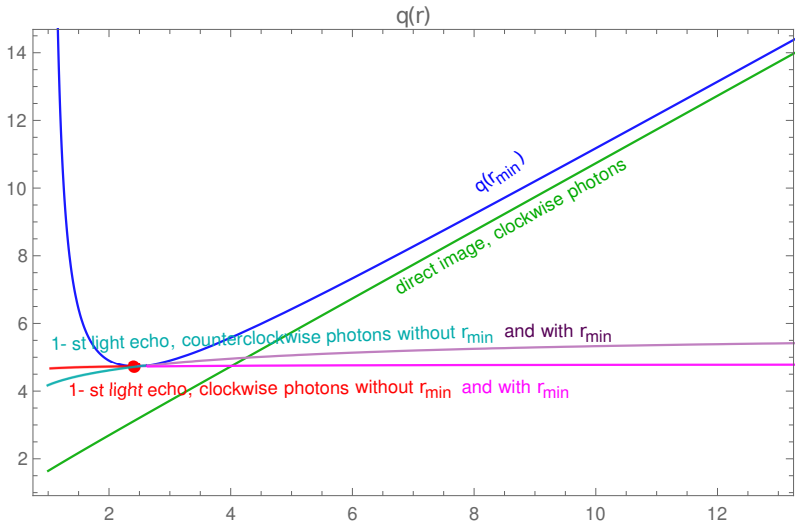}
	\caption{Relations $q(r)$ for photons producing the prime image (green curve) and the 1-st light echoes. Clockwise photons are moving along the trajectories with the growing polar angle $\theta$, while counterclockwise photons are moving in the opposite direction. Relations $q(r)$ are shown in different colors for clockwise and counterclockwise photon trajectories, with and without $r_{\rm min}$ along the photon trajectories. Red dot corresponds to the spherical photon orbit with $\lambda=0$. Blue curve $q(r_{\rm min})$ corresponds to the radial turning points $r=r_{\rm min}$, defined from equation $R(r_{\rm min})=0$.} 	\label{fig4}
\end{figure}

\begin{figure}
	\includegraphics[angle=0,width=0.65\textwidth]{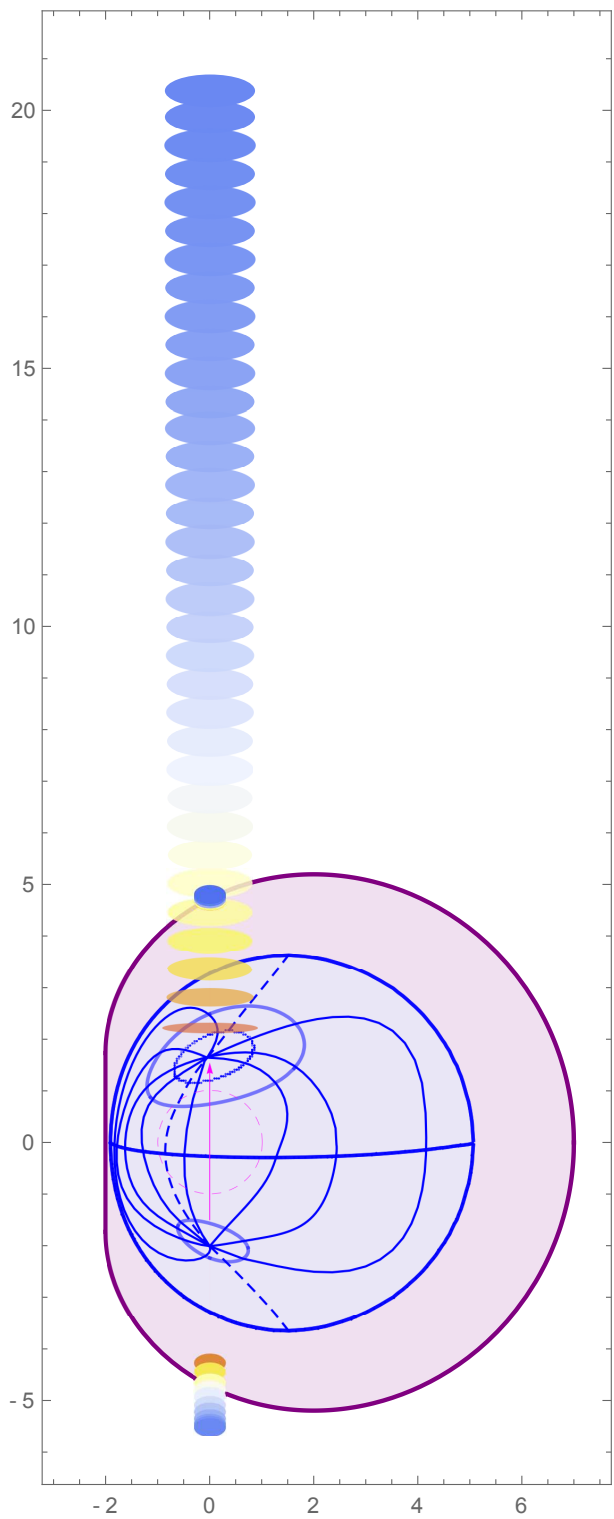}
	\caption{Direct image and 1-st light echoes of the outward moving hot spot in the jet from SgrA*  (supposing $a=1$) in discrete time intervals. Light echoes (secondary images) are viewed near the black hole shadow boundary (closed purple curve). The magenta arrow is the direction of the black hole rotation axis. The light blue region is the reconstructed image of event horizon globe \cite{doknaz19b,dokuch19,doknaz20,doknaz20b}.} \label{fig5}
\end{figure}

\begin{figure}
	\includegraphics[angle=0,width=0.5\textwidth]{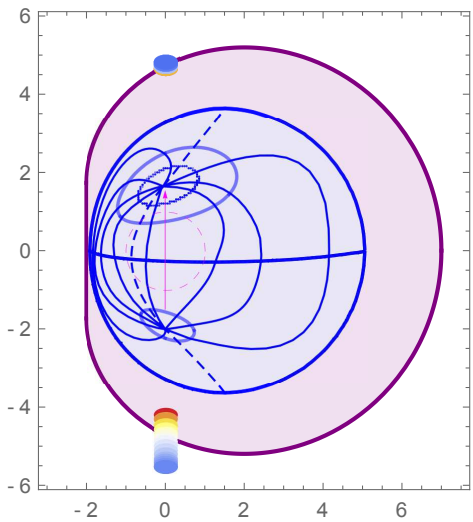}
	\caption{The 1-st light echoes of the outward moving hot spot in the jet from extreme Kerr black hole ($a=1$) SgrA*. They are concentrated near the black hole shadow boundary (closed purple curve). The light blue region is the reconstructed image of the northern hemisphere on the event horizon globe \cite{doknaz19b,dokuch19,doknaz20,doknaz20b}.} \label{fig6}
\end{figure}

\begin{figure}
	\includegraphics[angle=0,width=0.75\textwidth]{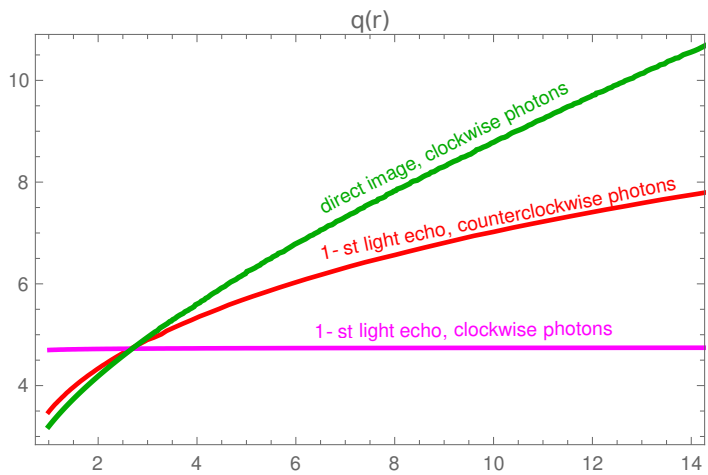}
	\caption{Relations $q(r)$ for photons, producing the prime image and the 1-st light echoes of the moving hot spot in  the jet from M87*. Clockwise photons (magenta curve) move along the trajectories with the growing polar angle $\theta$, while counterclockwise photons (red curve) move in the opposite direction.} 	\label{fig7}
\end{figure}

\begin{figure}
	\includegraphics[angle=0,width=0.8\textwidth]{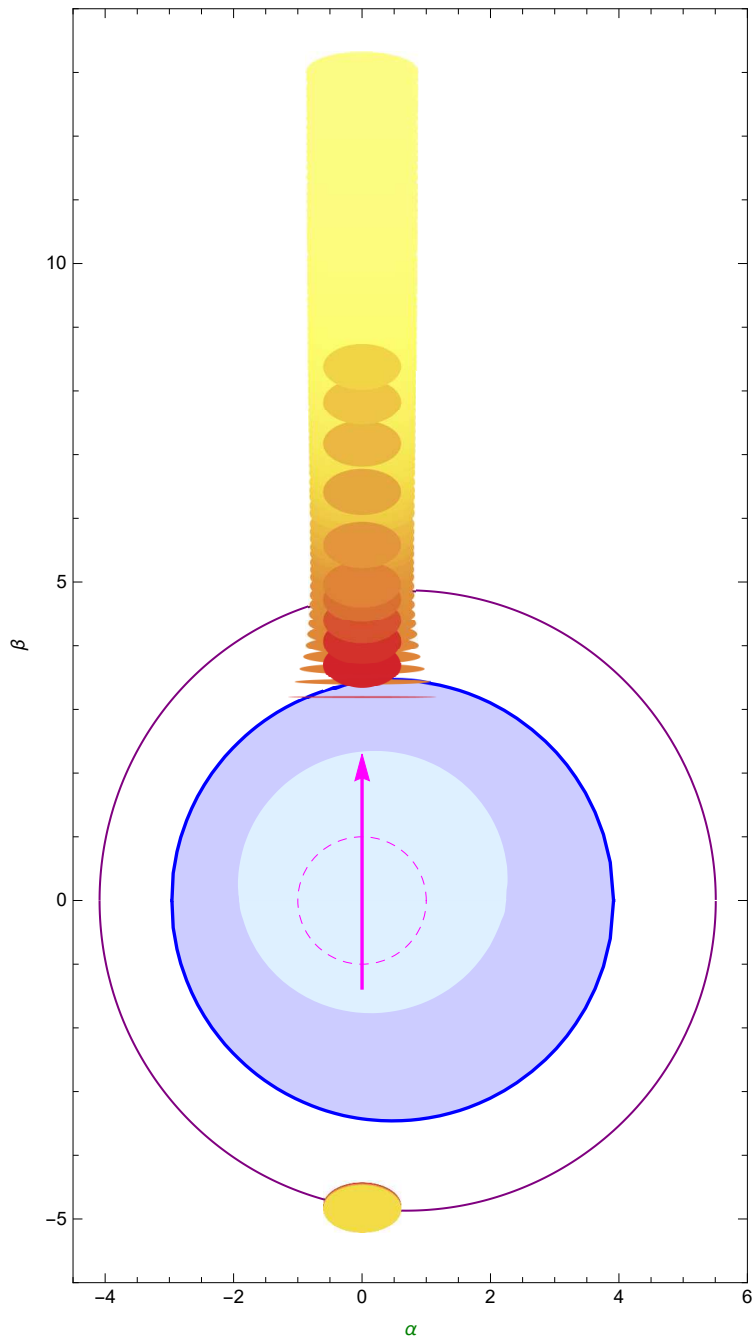}
	\hfill		
	\caption{The direct image and 1-st light echoes of the outward moving hot spot in the jet along the rotation axes of M87* in discrete time intervals. The luminous blob is starting at the radius $r=1.01r_{\rm h}$ (a little bit above the north pole of the event horizon globe) and finishing at the radius $r=13$. The colors of lensed images are related to the blob's black-body temperature which is supposed to be constant during the blob motion in the jet. The elliptic deformation of the lensed prime image of the small spherical blob in the strong gravitational field of extreme ($a=1$) Kerr black hole is shown. Light echoes (secondary images) are concentrated near the black hole shadow boundary (closed purple curve). The blue region is the reconstructed image of the event horizon globe. The light blue region is the reconstructed image of the northern hemisphere on the event horizon globe \cite{doknaz19b,dokuch19,doknaz20,doknaz20b}.}
	\label{fig8}
\end{figure}

\begin{figure}
	\includegraphics[angle=0,width=0.75\textwidth]{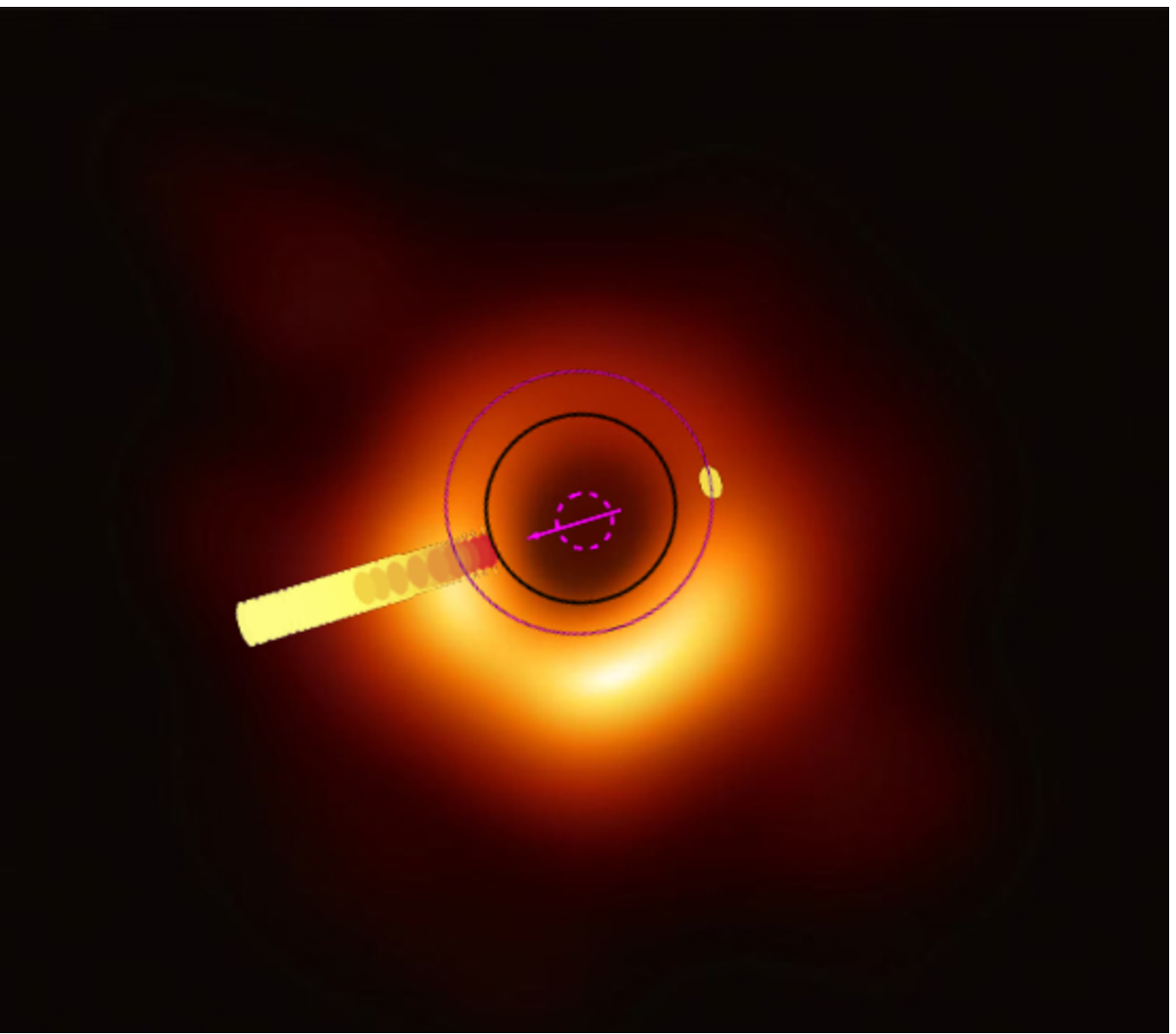}
	\caption{A composition of the Event Horizon Telescope image of M87* \cite{EHT1,EHT2,EHT3,EHT4,EHT5,EHT6} with the described numerical model of both the direct images and also the 1-st light echoes of the outward moving hot spot in the jet along the rotation axes of  M87* (shown in the discrete time intervals). The closed purple curve is the outline of the black hole shadow. The closed black curve is the outline of the event horizon image. The magenta arrow is the direction of the black hole rotation axis. The magenta dashed circle is a position of the black hole event horizon with radius $r_{\rm h}=1$ in the Euclidean space without gravity.} \label{fig9}
\end{figure}

Fig.~\ref{fig2} shows the corresponding $2D$ trajectories in the plane ($r,\theta$) of the same two photons as in Fig.~\ref{fig1}. 

See in Fig.~\ref{fig3} the similar $3D$ trajectories of photons, starting at the radius $r=1.49$ on the rotation axis of the black hole, and providing the prime image (green curve) and the 1-st light echo (red curve) of the outward moving hot spot. The orbital parameters $\lambda$ and $q=2.175$ for the prime image and, respectively,   $\lambda$ and $q=4.436$ for the 1-st light echo are derived from the numerical solutions of the integral equations of motion (\ref{eq24a}) and (\ref{eq24b}). The photon trajectory of this 1-st light echo is without the turning point $r_{\rm min}$, and this photon is moving counterclockwise in the $(r,\theta)$ plane. 

Fig.~\ref{fig4} demonstrates the corresponding relations for photons $q(r)$ for the prime image (green curve) and the 1-st light echoes. Clockwise photons are moving along the trajectories with the growing polar angle $\theta$, while counterclockwise photons are moving in the opposite direction. In this figure, the relations $q(r)$ are shown in different colors for clockwise and counterclockwise photon trajectories, with and without $r_{\rm min}$ along the photon trajectories. The red dot in this Figure corresponds to the spherical photon orbit $r=const$ with $\lambda=0$. Blue curve $q(r_{\rm min})$ corresponds to the radial turning points $r=r_{\rm min}$, defined from equation $R(r_{\rm min})=0$.

Fig.~\ref{fig5} shows both the direct images and the 1-st light echoes of the outward moving hot spot in the jet from SgrA* (supposing $a=1$) in discrete time intervals. The luminous blob is starting at the radius $r=1.01r_{\rm h}$ (a little bit above from the north pole of the event horizon globe) and finishing at the radius $r=20$. The colors of lensed images are related with the local black-body temperature of the blob, which is supposed to be constant during jet's blob motion. It is shown the elliptic deformation of the lensed prime image of the small spherical blob in the strong gravitational field of extreme ($a=1$) Kerr black hole. Light echoes (secondary images) are concentrated near the black hole shadow boundary (closed purple curve) on the celestial sphere. The images of all light echoes in this Figure and on the successive similar ones are artificially significantly increased to be comparable in brightness and sizes with prime images. The magenta dashed circle is the black hole event horizon with radius $r_{\rm h}=1$ in the Euclidean space without gravity. The light blue region is the reconstructed image of the hole event horizon globe \cite{doknaz19b,dokuch19,doknaz20,doknaz20b}. 

Fig.~\ref{fig6} shows the 1-st light echoes of the hot spot in the jet from the SgrA* (supposing $a=1$) in discrete time intervals moving clockwise and counterclockwise in the ($r,\theta$) plane. These light echoes are viewed near the projected position of the black hole shadow.

\section{Moving hot spot in the jet from M87*}

The supermassive black hole M87* is in the center of the galaxy M87. The famous radio and optical jet from M87* has a position angle PA$\,=288^\circ$ \cite{Walker18,Nalewajko20}. The corresponding viewing angle between the jet axis and line-of-sight is around $\theta_0=17^\circ$.

Fig.~\ref{fig7} is similar to Fig.~\ref{fig4} and demonstrates the corresponding relations for photons $q(r)$ for the prime image (green curve) and the 1-st light echoes of the moving hot spot (or massive blob) in the jet from M87* (by supposing $a=1$). In particular, the magenta curve corresponds to the clockwise photon trajectories, while the red curve corresponds to the counterclockwise ones. Clockwise photons are moving along the trajectories with the growing polar angle $\theta$, while counterclockwise photons are moving in the opposite direction. 

Meanwhile, Fig.~\ref{fig8} is similar to Fig.~\ref{fig5} and shows both the direct image and 1-st light echoes of the outward moving hot spot in the jet from M87* in discrete time intervals. The luminous blob is starting at the radius $r=1.01r_{\rm h}$ (a little bit above the north pole of the event horizon globe) and finishing at the radius $r=13$. The colors of lensed images are related to the blob's black-body temperature which is supposed to be constant during the blob motion in the jet. The elliptic deformation of the lensed prime image of the small spherical blob in the strong gravitational field of extreme Kerr black hole is shown. Light echoes (secondary images) are concentrated near the black hole shadow boundary (closed purple curve). The blue region is the reconstructed image of the event horizon globe. The light blue region is the reconstructed image of the northern hemisphere on the event horizon globe \cite{doknaz19b,dokuch19,doknaz20,doknaz20b}.

At last, Fig.~\ref{fig9} is a composition of the Event Horizon Telescope image of M87* \cite{EHT1,EHT2,EHT3,EHT4,EHT5,EHT6} with the described numerical model of both the direct images and also the 1-st light echoes of the outward moving hot spot in the jet from M87* (shown in the discrete time intervals). The closed purple curve is the outline of the black hole shadow. The closed black curve is the outline of the event horizon image. The magenta arrow is the direction of the black hole rotation axis. The magenta dashed circle is a position of the black hole event horizon with radius $r_{\rm h}=1$ in the Euclidean space without gravity.

\section{Discussion}

The lensing of any luminous object in the black hole's strong gravitational field leads to an infinite number of images. We numerically calculated the positions, forms and brightness for the   direct images and for the first light echoes of the bright spot moving radially outward in the jets from black holes SgrA* and M87*. 

The moving luminous blob is starting a little bit above from the north pole of the event horizon globe. The images of all light echoes in our Figures are artificially greatly enhanced to be comparable in brightness and sizes with the prime images. See the corresponding numerical animation of the lensed images of the small luminous blob moving along jet from the supermassive black holes SgrA* and M87* in \cite{doknaz20c}. Observations of these relativistic bright blobs by the future radio, optical and x-ray space-based interferometric observatories can provide the experimental verification of the general relativity and its modifications in the strong field limit. The most evident direction in this verification is a precise measurement the  black hole shadows for the following comparison with predictions of general relativity and modified gravity theories.

In this paper we investigate the gravitational lensing of the moving bright spot in the jet along a rotational axis of the Kerr black hole in the framework of general relativity (by calculating photon trajectories and  arrival times of photons to a distant observer). Quite the similar investigations may be done by using the corresponding black hole solutions in the numerous general relativity modifications (with photon trajectories and arrival times of photons to a distant observer different from the prediction of general relativity). 

In the nearest future it would be possible to choose the preferable gravity theory by using observational data from the future advanced telescopes including the laser gravitational wave interferometer antennas and space-based sub-millimeter interferometric observatories. The current one and most promising at the moment is EHT with angular resolution of a few gravitational radii of M87 and SgrA. But several new telescopes are coming into observations soon. In the optical infrared band, James Webb Space Telescope (2021), Giant Magellan Telescope (2023), Extremely Large Telescope (2024), Thirty Meter Telescope (2027) with angular resolution of ~100 mas  should be able to resolve galactic jets and define the gravitational potential. The Cerenkov Telescope Array (CTA) is scheduled to start collecting data in 2022. It will have ten times the sensitivity of existing telescopes  should greatly improve our understanding of jet beaming and variability and help us determine the geometry, speeds and emission mechanisms of black hole jets. In the X-ray band, the Imaging X-ray Polarimetry Explorer (IXPE, 2021) should be able to determine the geometry and the emission mechanism of active galactic nuclei and ATHENA (early 2030s) will improve on existing telescopes (like ALMA, ELT, JWST, SKA, CTA, etc) in sensitivity and resolution.

\begin{acknowledgements}

This work was supported in part by the Russian Foundation for Basic Research grant 18-52-15001a.

\end{acknowledgements}

\appendix

\section{Appendix}
\label{Appendix}

We describe here some mathematical details and physical approaches for our numerical calculations of the bright spot motion along the black hole rotation axes.

\subsection{Locally Non-Rotating Frames}

The most physically appropriate coordinate frame for Kerr black hole is a so-called  Locally Non-Rotating Frames (LNRF) \cite{Bardeen70,BPT}. The orthonormal tetrad 
\begin{equation}
	{\mathbf e}_{(i)}=e^\mu_{\phantom{1}(i)}\frac{\partial}{\partial x^\mu}, 
	\qquad 
	{\mathbf e}^{(i)}=e_\mu^{\phantom{1}(i)}{\mathbf d}x^\mu
	\label{basis}
\end{equation}
relates the Boyer--Lindquist coordinates $(t,r,\theta,\phi)$ with the similar ones for physical observers in the LNRF:
\begin{eqnarray}
	{\mathbf e}_{(t)}&\!=\!&e^{-\nu}\left(\frac{\partial}{\partial t}
	\!+\!\omega\frac{\partial}{\partial\varphi}\right)\!=\!
	\left(\frac{A}{\Sigma\Delta}\right)^{1/2}\!\frac{\partial}{\partial t}\!+\!
	\frac{2Mat}{(A\Sigma\Delta)^{1/2}}\frac{\partial}{\partial\varphi}, \label{et} 
	\nonumber \\
	{\mathbf e}_{(r)}&=&e^{-\mu_1}\frac{\partial}{\partial r} =
	\left(\frac{\Delta}{\Sigma}\right)^{1/2}\frac{\partial}{\partial r}, \label{er} 
	\nonumber \\
	{\mathbf e}_{(\theta)}&=&e^{-\mu_2}\frac{\partial}{\partial\theta} =
	\frac{1}{\Sigma^{1/2}}\frac{\partial}{\partial\theta}, \label{etheta} 
	\nonumber \\
	{\mathbf e}_{(\varphi)}&=&e^{-\psi}\frac{\partial}{\partial\varphi} =
	\left(\frac{\Sigma}{A}\right)^{1/2}\!\frac{1}{\sin\theta}
	\frac{\partial}{\partial\varphi}.\label{evarphi} 
\end{eqnarray}
The related basis differential 1-forms for LNRF are
\begin{eqnarray}
	{\mathbf e}^{(t)}&=&e^{\nu}{\mathbf d}t=
	\left(\frac{\Sigma\Delta}{A}\right)^{1/2}\!{\mathbf d}t, \label{et2} \\
	{\mathbf e}^{(r)}&=&e^{\mu_1}{\mathbf d}t=
	\left(\frac{\Sigma}{\Delta}\right)^{1/2}\!{\mathbf d}r, \label{er2}  \\
	{\mathbf e}^{(\theta)}&=&e^{\mu_2}{\mathbf d}t=
	\Sigma^{1/2}\,{\mathbf d}\theta, \label{etheta2}  \\
	{\mathbf e}^{(\varphi)}&=&-\omega e^{\psi}{\mathbf d}t+e^{\psi}{\mathbf d}\varphi
	=-\frac{2Mar\sin\theta}{(\Sigma A)^{1/2}}\,{\mathbf d}t+
	\left(\frac{A}{\Sigma}\right)^{1/2}\!\!\sin\theta\,{\mathbf d}\varphi. \label{evarphi2} 
\end{eqnarray}

\subsection{Equations of motion for test particles}

Brandon Carter \cite{Chandra,Carter68,BPT,mtw} derived the remarkable first order differential equations of motion in the Kerr space-time, which meant as for the analytical and also for numeric calculations of the test particle trajectories. These trajectories depend on the integrals of motion: $\mu$ --- the test particle mass, $E$ --- the test particle total energy, $L$ --- test particle azimuth angular momentum, and the very specific Carter constant $Q$, defining the non-equatorial motion of the test particle. The motion of test particles is bounded to an equatorial plane of the metric if $Q = 0$. 
\begin{eqnarray} 
	\Sigma\frac{dr}{d\tau} &=& \pm \sqrt{R(r)}, 
	\label{rmot} \\
	\Sigma\frac{d\theta}{d\tau} &=& \pm\sqrt{\Theta(\theta)}, \label{thetamot} \\
	\Sigma\frac{d\phi}{d\tau} &=& L\sin^{-2}\theta+a(\Delta^{-1}P-E), 
	\label{varphiamot}	\\
	\Sigma\frac{dt}{d\tau} &=& a(L-aE\sin^{2}\theta)+(r^2+a^2)\Delta^{-1}P,
	\label{tmot}	
\end{eqnarray}
where $\tau$ is a proper time of the massive ($\mu\neq0$) test particles or a corresponding affine parametrization for the massless ($\mu=0$) test particles. The radial potential $R(r)$ in these equations, which governs the radial motion of test particles, is
\begin{equation}	\label{Rr} 
	R(r) = P^2-\Delta[\mu^2r^2+(L-aE)^2+Q],
\end{equation}
where 
\begin{equation}	\label{P} 
	P=E(r^2+a^2)-a L.
\end{equation}
Respectively, the polar potential $\Theta(\theta)$ is
\begin{equation}	\label{Vtheta} 
	\Theta(\theta) = Q-\cos^2\theta[a^2(\mu^2-E^2)+L^2\sin^{-2}\theta].
\end{equation}
Note that zeros of these potentials define the radial and polar turning points $dR/d\tau=0$ and $d\Theta/d\tau=0$, correspondingly. 

It is useful to define the orbital parameters for the massive test particles, $\gamma=E/\mu$, $\lambda=L/E$ and $q^2=Q/E^2$. Notice that there are also possible particle trajectories with $Q<0$, which do not reach the space infinity and, consequently, a distant observer (formally at $r=\infty$). We will not consider such trajectories in this article.

\subsection{Integral equations for test particle motion}

In our numerical calculations, we also use the integral equations for test particle motion (\ref{rmot})--(\ref{tmot}):
\begin{equation}\label{eq2425a}
	\quad \:\:\:	\fint\frac{dr}{\sqrt{R(r)}}
	=\fint\frac{d\theta}{\sqrt{\Theta(\theta)}}, 
\end{equation}
\begin{equation}\label{eq2425b}
	\tau=\fint\frac{r^2}{\sqrt{R(r)}}\,dr
	+\fint\frac{a^2\cos^2\theta}{\sqrt{\Theta(\theta)}}\,d\theta,
\end{equation}
\begin{equation}\label{eq25ttc}
	\phi=\fint\frac{aP}{\Delta\sqrt{R(r)}}\,dr
	+\fint\frac{L-aE\sin^2\theta}{\sin^2\theta\sqrt{\Theta(\theta)}}\,d\theta, 
\end{equation}
\begin{equation}\label{eq25ttd}
	t=\fint\frac{(r^2+a^2)P}{\Delta\sqrt{R(r)}}\,dr
	+\fint\frac{(L-aE\sin^2\theta)a}{\sqrt{\Theta(\theta)}}\,d\theta. 
\end{equation}
The integrals in (\ref{eq2425a})--(\ref{eq25ttd}) are the line (or curve) integrals monotonically growing along the test particle trajectories. For example, the line integrals in (\ref{eq2425a}) add up to the ordinary ones in the absence of both the radial and polar turning points on the particle test particle trajectory:
\begin{equation}\label{eq24a}
	\int^{r_s}_{r_0}\frac{dr}{\sqrt{R(r)}}=  \int_{\theta_0}^{\theta_s}\frac{d\theta}{\sqrt{\Theta(\theta)}}.
\end{equation}
Here $r_s$ and $\theta_s$ are the initial test particle (e.\,g., photon) radial and polar  coordinates, while $r_0\gg r_{\rm h}$ and $\theta_0$ is the corresponding final (finishing) points on the trajectory  (e.\,g., the photon detection point by a distant telescope). The second example is a case when there is only one turning point in the polar direction, $\theta_{\rm min}(\lambda,q)$ (derived from the equation $\Theta(\theta)=0$). The corresponding line integrals in (\ref{eq2425a}) add up now to the ordinary ones: 
\begin{equation}\label{eq24b}
	\int_{r_s}^{r_0}\frac{dr}{\sqrt{R(r)}}
	=\int_{\theta_{\rm min}}^{\theta_s}\frac{d\theta}{\sqrt{\Theta(\theta)}}
	+\int_{\theta_{\rm min}}^{\theta_0}\frac{d\theta}{\sqrt{\Theta(\theta)}}.
\end{equation}
The most complicated case, which we consider in this paper, corresponds to the test particle trajectory with the one turning point in the polar direction, $\theta_{\rm min}(\lambda,q)$ (derived from the equation $\Theta(\theta)=0$), and the one turning point in the radial direction, $r_{\rm min}(\lambda,q)$ (derived from the equation $R(r)=0$). The corresponding line integrals in (\ref{eq2425a}) in the our most complicated case add up to the following ordinary ones: 
\begin{equation}\label{eq24c}
	\int_{r_{\rm min}}^{r_s}\!\!\frac{dr}{\sqrt{R(r)}}
	+\int_{r_{\rm min}}^{r_0}\!\!\frac{dr}{\sqrt{R(r)}}
	=\!\!\int_{\theta_{\rm min}}^{\theta_s}\!\!\frac{d\theta}{\sqrt{\Theta(\theta)}}
	+\!\!\int_{\theta_{\rm min}}^{\theta_0}\!\!\frac{d\theta}{\sqrt{\Theta(\theta)}}.
\end{equation}
It is clear that integral equations (\ref{eq2425a})--(\ref{eq25ttd}) for test particle trajectories with more numbers of turning points add up to the ordinary integrals in similar ways.

\subsection{Energy shift of photons emitted by the moving bright spot}

We suppose that a bright spherical massive blob of hot plasma with a mass $\mu$ and total energy $E=\mu$ (parabolic motion) is moving away from the extreme Kerr black hole ($a=1$) with ballistic velocity along the black hole rotation axis, starting very close to the event horizon radius, $r_{\rm h}=1$. To disregard the tidal effects, we additionally suppose that the radius of bright massive blob, $r_{\rm b}$, is negligible in comparison with the event horizon radius $r_{\rm b}\ll r_{\rm h}$.  It is also supposed that the radius of the blob remains constant during the motion.

From equation (\ref{Vtheta}) for the polar potential  $\Theta(\theta)$ it follows that all photons reaching a distant observer and starting from the black hole rotation axis (at $\theta_s=0$ and $r\geq r_{\rm h}$) must have orbital parameter $\lambda=L/E=0$. So, our task is reduced to finding only one orbital parameter, $q(r)$ for these photons. 

Orbital parameters $\lambda=L/E$ and $q=\sqrt{Q}/E$ of the photon trajectory, reaching a distant telescope (placed at the radius $r_0>>r_{\rm h}$, at the polar angle $\theta_0$ and at the azimuth angle $\varphi_0$), are directly related with the corresponding  impact parameters, viewed at the celestial sphere \cite{CunnBardeen72,CunnBardeen73,Bardeen73}: 
\begin{equation}
	\alpha =-\frac{\lambda}{\sin\theta_0}, \quad
	\beta = \pm\sqrt{\Theta(\theta_0)},
	\label{alphabeta} 
\end{equation}
where $\Theta(\theta)$ defined in equation (\ref{Vtheta}). The impact parameters $\alpha$ and $\beta$ are called, respectively, the horizontal and vertical impact parameters.

We also take into account the gravitational red-shift and Doppler effect of photons emitted by the luminous blob of plasma moving along the jet and detected by a distant observer. The orthonormal Locally Non-Rotating Frames (LNRF) from equations (\ref{basis})--(\ref{evarphi2}) is suitable for calculations of the corresponding energy shift of these photons and energy flux from the bright blob detected by a distant observer.  

A radial velocity component of the luminous blob with a mass $\mu$, moving along the jet with the azimuth angular momentum $L=0$ in the LNRF is 
\begin{equation}\label{Vr}
	V\equiv V^{(r)}=\frac{u^\mu e^{(r)}_\mu}{u^\nu e^{(t)}_\nu} 
	=\frac{\sqrt{R(r)}}{(r^2+a^2)\gamma}.
\end{equation}
Here, $\gamma=E/\mu$ is the Lorentz gamma-factor, $u^\mu=dx^\mu/ds$ is the 4-velocity of the blob, defined by differential equations (\ref{rmot})--(\ref{tmot}), and $R(r)$ is the radial potential from equation~(\ref{Rr}) with the parameter $L=0$. 

All photons, reaching a distant observer, start from the moving blob with the orbital parameter $\lambda=0$. In this case, the corresponding components of the photon 4-momentum $p^{(\mu)}$ in the LNRF are 
\begin{eqnarray}
	p^{(\varphi)}&=&g^{\mu\nu}p_\nu e_\mu^{(\varphi)} =0, 
	\label{pphi} \\
	p^{(t)}&=&g^{\mu\nu}p_\nu e_\mu^{(t)} 
	=\sqrt{\frac{r^2+a^2}{\Delta}}, 
	\label{pt} \\
	p^{(r)}&=&g^{\mu\nu}p_\nu e_\mu^{(r)}
	=\sqrt{\frac{r^2+a^2}{\Delta}-\frac{r^2+q^2}{r^2+a^2}},
	\label{pr}
\end{eqnarray}
where $\Delta$ defined in equation (\ref{Sigma}). The condition $p^{(i)}p_{(i)}=0$ determines the fourth component of the photon 4-momentum. The photon energy in the LNRF is $E_{\rm LNRF}=p^{(t)}$. Meantime, the corresponding photon energy in the comoving frame of the massive blob moving with a radial velocity $V$ relative to the LNRF is 
\begin{equation}\label{E}
	{\cal{E}}(r,q)=\frac{p^{(t)}-Vp^{(r)}}{\sqrt{1-V^2}}.
\end{equation}
As a result, the photon energy shift (ratio of the photon frequency detected by a distant observer to the frequency of the same photon in the comoving frame of the blob) is $g(r,q)=1/{\cal{E}}(r,q)$. 
This energy takes into account gravitational red-shift and the Doppler effect.

\subsection{Energy flux from the moving bright spot}

We incorporate the photon energy shift $g(r,q)$ from equation (\ref{E}) into the very useful Cunningham--Bardeen formalism \cite{CunnBardeen72,CunnBardeen73} for numerical calculation of the energy flux detected by a distant observer from the moving blob of finite size. We also numerically calculate the elliptic deformation of the lensed prime image of the small spherical blob in the strong gravitational field of extreme Kerr black hole (see more details of these calculations in \cite{doknaz17,doknazsm19,doknaz19,doknaz19b,dokuch19,doknaz20,doknaz20b})

The corresponding flux of energy detected by a distant observer from the bright spot in the jet is the double integral of the surface brightness over the angular spread of the bright spot image \cite{CunnBardeen73}:
\begin{equation}\label{F0}
	F_0=\iint\frac{d\alpha d\beta}{r_0^2} \frac{L}{4\pi^2b^2}g(r,q)^2.
\end{equation}
In this equation $\alpha$ and $\beta$ are respectively the horizontal and vertical impact parameters from (\ref{alphabeta}), $r_0$ is a distance to the bright spot, $b$ is a radius of the bright spot (bright spherical massive blob of hot plasma) and $g(r,q)$ is the detected photon energy shift from equation (\ref{E}). The double integral (\ref{F0}) in the case of a distant observer at $r_0\gg r_{\rm h}$ is well approximated in the deviation of the photon trajectory from the central one in the gravitational field of the black hole:
\begin{equation}\label{F0}
	F_0=\frac{L}{4\pi^2b^2r_0^2}g(r,q)^2\iint d\alpha d\beta.
\end{equation}
In the used Cunningham--Bardeen formalism it is developed the effective perturbation method for numerical calculation of the double integral in this equation by taking into account the intersection of photon trajectory with the plane of the bright spot disk perpendicular to the radial direction in the local rest frame the bright spot (see details of this effective perturbation method in \cite{CunnBardeen73}).

Results of all our numerical calculations are illustrated in Figs.~\ref{fig1}--\ref{fig9} and in numerical animation \cite{doknaz20c}.

\end{document}